\documentclass[11pt]{article}
\newcommand{\bq}{\begin{equation}}
\newcommand{\eq}{\end{equation}}
\newcommand{\ba}{\begin{eqnarray}}
\newcommand{\ea}{\end{eqnarray}}

\newcommand{\nl }{ \nonumber  }

\newcommand{\p}{\partial}
\newcommand{\h}{\hspace{.5cm}}

\newcommand{\la}{\lambda}
\newcommand{\La}{\Lambda}
\newcommand{\Di}{\left(\p_0-\la^{i}\p_i\right)}
\newcommand{\Dj}{\left(\p_0-\la^{j}\p_j\right)}

\begin{document}
\begin{center}
{\bf PROBE BRANES DYNAMICS:\\
EXACT SOLUTIONS IN GENERAL BACKGROUNDS
\vspace*{0.5cm}\\ P. Bozhilov}
\\ {\it Department of Theoretical and Applied Physics, \\
Shoumen University, 9712 Shoumen, Bulgaria\\
E-mail:} p.bozhilov@shu-bg.net\\
and\\
{\it The Abdus Salam International Centre for Theoretical Physics,
Trieste, Italy\\
E-mail:} bozhilov@ictp.trieste.it\\
\end{center}
\vspace*{0.5cm}

We consider probe $p$-branes and D$p$-branes dynamics in $D$-dimensional
string theory backgrounds of general type. Unified description for the
tensile and tensionless branes is used. We obtain exact solutions of their
equations of motion and constraints in static gauge as well as in more
general gauges. Their dynamics in the whole space-time is also analyzed and
exact solutions are found.

\vspace*{.5cm}
PACS number(s):11.25.-w; 11.27.+d; 11.30.-j

\vspace*{.5cm}

\section{Introduction}

The {\it probe} branes approach for studying issues in the string/M-theory
uses an approximation, in which one neglects the back-reaction of the branes on
the background. In this sense, the probe branes are multidimensional
dynamical systems, evolving in given, variable in general, external fields.

The probe branes method is widely used in the string/M-theory to
investigate many different problems at a classical, semiclassical and
quantum levels. The literature in this field of research can be {\it
conditionally} divided into several parts. One of them is devoted to the
properties of the probe branes themselves, e.g., \cite{CM98}-\cite{HHW02}.
The subject of another part of the papers is to probe the geometries of
the string/M-theory backgrounds, e.g., \cite{DKL94}-\cite{HJS02}. One
another part can be described as connected with the investigation of the
correspondence between the string/M-theory geometries and their field
theory duals, e.g., \cite{K99_1}-\cite{RV02}. Let us also mention the
application of the probe branes technique in the 'Mirage cosmology'-
approach to the brane world scenario, e.g., \cite{KK99}-\cite{TBS02}.

In view of the wide implementation of the probe branes as a tool for
investigation of different problems in the string/M-theory, it will be useful
to have a method describing their dynamics, which is general enough to include
as many cases of interest as possible, and on the other hand, to give the
possibility for obtaining {\it explicit exact} solutions.

In this article, we propose such an approach, which is appropriate for
$p$-branes and D$p$-branes, for arbitrary worldvolume and space-time
dimensions, for tensile and tensionless branes, for different variable
background fields with minimal restrictions on them, and finally, for
different space-time and worldvolume gauges (embeddings).

The paper is organized as follows. In Section 2, we perform an analysis in
order to choose the brane actions, which are most appropriate for our
purposes - do not contain square roots, generate only {\it independent}
constraints and give a unified description for tensile and tensionless
branes. In Section 3, we formulate the conditions, under which the probe
branes dynamics reduces to a particle-like one. Then, we investigate the
reduced dynamics for three different types of brane embeddings, starting
with the usually used static gauge one. In the last part of this section,
we find explicit exact solutions of the branes equations of motion. In
Section 4, we discuss the obtained results.

\setcounter{equation}{0}
\section{Actions}

Before considering the problem for obtaining exact brane solutions in
general string theory backgrounds, it will be useful first to choose
appropriate actions, which  will facilitate our task. Generally speaking,
there are two types of brane actions - with and without square roots
\footnote{Examples of these two type of actions are the Nambu-Goto and
Polyakov actions for the string.}. The former ones are not well suited to
our purposes, because the square root introduces additional nonlinearities
in the equations of motion. Nevertheless, they have been used when
searching for exact brane solutions in fixed backgrounds, because there
are no constraints in the Lagrangian description and one has to solve only
the equations of motion. The other type of actions contain additional
worldvolume fields (Lagrange multipliers). Varying with respect to them,
one obtains constraints, which, in general, are not independent. Starting
with an action without square root, one escapes the nonlinearities
connected with the square root, but has to solve the equations of motion
and the (dependent) constraints.

Independently of their type, all actions proportional to the brane tension
cannot describe the tensionless branes. The latter appear in many important
cases in the string theory, and it is preferable to have a unified
description for tensile and tensionless branes.

Our aim in this section is to find brane actions, which do not
contain square roots, generate only independent constraints and give a unified
description for tensile and tensionless branes.

\subsection{$P$-brane actions}

The Polyakov type action for the bosonic $p$-brane in a $D$-dimensional
curved space-time with metric tensor $g_{MN}(x)$, interacting with a
background (p+1)-form gauge field $b_{p+1}$ via Wess-Zumino term, can be
written as \ba\label{pa} S_p^{P}=&-&\int
d^{p+1}\xi\Bigl\{\frac{T_p}{2}\sqrt{-\gamma}\left[\gamma^{mn} \p_m X^M\p_n
X^N g_{MN}(X)-(p-1)\right]\\ \nl &-&Q_p\frac{ \varepsilon^{m_1\ldots
m_{p+1}}}{(p+1)!} \p_{m_1}X^{M_1}\ldots\p_{m_{p+1}}X^{M_{p+1}}
b_{M_1\ldots M_{p+1}}(X)\Bigr\},\\ \nl && \p_m=\p/\p\xi^m,\h m,n =
0,1,\ldots,p;\h M,N = 0,1,\ldots,D-1,\ea where $\gamma$ is the determinant
of the auxiliary worldvolume metric $\gamma_{mn}$, and  $\gamma^{mn}$ is
its inverse. The position of the brane in the background space-time is
given by $x^M=X^M(\xi^m)$, and $T_p$, $Q_p$ are the $p$-brane tension and
charge, respectively. If we consider the action (\ref{pa}) as a bosonic
part of a supersymmetric one, we have to set $Q_p=\pm T_p$. In what
follows, $Q_p = T_p$.

The requirement that the variation of the action (\ref{pa}) with respect
to $\gamma_{mn}$ vanishes, leads to \ba\label{dc}
\left(\gamma^{kl}\gamma^{mn}-2\gamma^{km}\gamma^{ln}\right)G_{mn}
=(p-1)\gamma^{kl},\ea where $G_{mn}=\p_m X^M\p_n X^N g_{MN}(X)$ is the
metric induced on the $p$-brane worldvolume. Taking the trace of the above
equality, one obtains \ba\nl \gamma^{mn}G_{mn}=p+1,\ea i.e., $\gamma^{mn}$
is the inverse of $G_{mn}$: $\gamma^{mn}=G^{mn}$. If one inserts this back
into (\ref{pa}), the result will be the corresponding Nambu-Goto type
action $(G\equiv\det(G_{mn}))$: \ba\label{nga} S_p^{NG}&=& \int
d^{p+1}\xi\mathcal{L}^{NG}\\ \nl &=& - T_p\int d^{p+1}\xi\left[
\sqrt{-G}-\frac{ \varepsilon^{m_1\ldots m_{p+1}}}{(p+1)!}
\p_{m_1}X^{M_1}\ldots\p_{m_{p+1}}X^{M_{p+1}} b_{M_1\ldots
M_{p+1}}(X)\right].\ea This means that the two actions, (\ref{pa}) and
(\ref{nga}), are classically equivalent.

As already discussed, the action (\ref{nga}) contains a square root,
the constraints (\ref{dc}), following from (\ref{pa}), are not independent
and none of these actions is appropriate for description of the {\it
tensionless} branes. To find an action of the type we are looking for,
we first compute the
explicit expressions for the generalized momenta, following from
(\ref{nga}): \ba\nl P_M(\xi)=-T_p\left(\sqrt{-G}G^{0n}\p_nX^Ng_{MN}
-\p_1X^{M_1}\ldots\p_pX^{M_p}b_{MM_1\ldots M_p}\right).\ea It can be
checked that $P_M(\xi)$ satisfy the constraints \ba\nl
&&\mathcal{C}_0\equiv g^{MN}P_MP_N -2T_pg^{MN}D_{M1\ldots p}P_N
+T^2_p\left[GG^{00}+(-1)^pD_{1\ldots p M}g^{MN}D_{N1\ldots p}\right]=0, \\
\nl &&\mathcal{C}_i\equiv P_M\p_iX^M=0,\h (i=1,\ldots,p),\ea where we have
introduced the notation \ba\nl D_{M1\ldots p}\equiv b_{MM_1\ldots
M_p}\p_1X^{M_1}\ldots\p_pX^{M_p}.\ea

Let us now find the canonical Hamiltonian for this dynamical system. The
result is: \ba\nl H_{canon} = \int d^p\xi\left(P_M\p_0X^M
-\mathcal{L}^{NG}\right)= 0.\ea Therefore, according to Dirac \cite{D64},
we have to take as a Hamiltonian the linear combination of the first class
primary constraints $\mathcal{C}_n$: \footnote{In the case under
consideration, secondary constraints do not appear. The first class
property of $\mathcal{C}_n$ follows from their Poisson bracket algebra.}
\ba\nl H = \int d^p\xi\mathcal{H} = \int d^p\xi
\left(\lambda^0\mathcal{C}_0 + \lambda^i\mathcal{C}_i\right).\ea The
corresponding Hamiltonian equations of motion for $X^M$ are \ba\nl
\left(\p_0-\lambda^i\p_i\right)X^M =
2\lambda^0g^{MN}\left(P_N-T_pD_{N1\ldots p}\right),\ea from where one
obtains the explicit expressions for $P_M$ \ba\label{nm} P_M
=\frac{1}{2\lambda^0}g_{MN}\left(\p_0-\lambda^j\p_j\right)X^N
+T_pD_{M1\ldots p}.\ea With the help of (\ref{nm}), one arrives at the
following configuration space action \ba\label{oa} S_p&=&\int
d^{p+1}\xi\mathcal{L}_p=\int d^{p+1}\xi\left(P_M\p_0X^M
-\mathcal{H}\right)\\ \nl &=& \int
d^{p+1}\xi\Bigl\{\frac{1}{4\lambda^0}\Bigl[g_{MN}\left(X\right)
\left(\p_0-\lambda^i\p_i\right) X^M\left(\p_0-\lambda^j\p_j\right)X^N
-\left(2\lambda^0T_p\right)^2GG^{00}\Bigr]\\ \nl &+&T_p b_{M_0\ldots
M_p}(X)\p_0X^{M_0}\ldots\p_pX^{M_p} \Bigr\}\\ \nl &=&\int
d^{p+1}\xi\Bigl\{\frac{1}{4\lambda^0}\Bigl[
G_{00}-2\lambda^{j}G_{0j}+\lambda^{i}\lambda^{j}G_{ij}
-\left(2\lambda^0T_p\right)^2GG^{00}\Bigr]\\ \nl &+&T_p b_{M_0\ldots
M_p}(X)\p_0X^{M_0}\ldots\p_pX^{M_p} \Bigr\},\ea which does not contain
square root, generates the independent $(p+1)$ constraints, as we will
show below, and in which the limit $T_p\to 0$ may be taken. For other
actions, allowing for unified description of tensile and tensionless
$p$-branes, see \cite{BZ93} - \cite{HLU94}.

It can be proven that this action is classically equivalent to the previous
two actions. It is enough to show that (\ref{nga}) and (\ref{oa}) are
equivalent, because we already saw that this is true for (\ref{pa}) and
(\ref{nga}).

Varying the action $S_p$ with respect to Lagrange multipliers $\lambda^m$
and requiring these variations to vanish, one obtains the constraints
\ba\label{pbic0} &&G_{00}-2\lambda^{j}G_{0j}+\lambda^{i}\lambda^{j}G_{ij}
+\left(2\lambda^0T_p\right)^2GG^{00}=0,\\
\label{pbicj} &&G_{0j}-\lambda^{i}G_{ij}=0.\ea
By using them, the Lagrangian density $\mathcal{L}_p$ from (\ref{oa}) can be
rewritten in the form \ba\label{il} \mathcal{L}_p=-T_p
\sqrt{-GG^{00}\left[G_{00}-G_{0i}\left(G^{-1}\right)^{ij}G_{j0}\right]}
+T_p b_{M_0\ldots M_p}(X)\p_0X^{M_0}\ldots\p_pX^{M_p}.\ea Now, applying
the equalities \ba\label{sg} GG^{00}=\det\left(G_{ij}\right)\equiv\mathbf{G},
\h G=\left[G_{00}-G_{0i}\left(G^{-1}\right)^{ij}G_{j0}\right]\mathbf{G},\ea
one finds that \ba\nl
G^{00}\left[G_{00}-G_{0i}\left(G^{-1}\right)^{ij}G_{j0}\right]=1.\ea
Inserting this in (\ref{il}), one obtains the Nambu-Goto type Lagrangian
density $\mathcal{L}^{NG}$ from (\ref{nga}). Thus, the classical
equivalence of the actions (\ref{nga}) and (\ref{oa}) is established.

We will work further in the gauge $\lambda^m=constants$, in which the
equations of motion for $X^M$, following from (\ref{oa}), are given by
\ba\label{pbem} &&g_{LN}\left[\Di\Dj X^N - \left(2\lambda^0T_p\right)^2
\p_i\left(\mathbf{G}G^{ij}\p_j X^N\right)\right]\\ \nl
&&+\Gamma_{L,MN}\left[\Di X^M \Dj X^N - \left(2\lambda^0T_p\right)^2
\mathbf{G}G^{ij}\p_i X^M \p_j X^N\right]\\ \nl
&&=2\la^0 T_p H^{\mathbf{b}}_{LM_0\ldots M_{p}}\p_0 X^{M_0}\ldots
\p_p X^{M_p},\ea
where $\mathbf{G}$ is defined in (\ref{sg}),
\ba\nl \Gamma_{L,MN}=g_{LK}\Gamma^K_{MN}=\frac{1}{2}\left(\p_Mg_{NL}
+\p_Ng_{ML}-\p_Lg_{MN}\right)\ea
are the components of the symmetric connection compatible with the metric
$g_{MN}$ and $H^{\mathbf{b}}_{p+2}=db_{p+1}$ is the field strength of the
$(p+1)$-form gauge potential $b_{p+1}$.

\subsection{D$p$-brane actions}

The Dirac-Born-Infeld type action for the bosonic part of the super-
D$p$-brane in a $D$-dimensional space-time with metric tensor $g_{MN}(x)$,
interacting with a background (p+1)-form Ramond-Ramond gauge field
$c_{p+1}$ via Wess-Zumino term, can be written as \ba\label{dbia}
S^{DBI}&=&-T_{D_p}\int d^{p+1}\xi
\Bigl\{e^{-a(p,D)\Phi}\sqrt{-\det\left(G_{mn} + B_{mn} +
2\pi\alpha'F_{mn}\right)}\\ \nl &-&\frac{ \varepsilon^{m_1\ldots
m_{p+1}}}{(p+1)!} \p_{m_1}X^{M_1}\ldots\p_{m_{p+1}}X^{M_{p+1}}
c_{M_1\ldots M_{p+1}}\Bigr\}.\ea $T_{D_P}$=$(2\pi)^{-(p-1)/2}g_s^{-1}T_p$
is the D-brane tension, $g_s$ = $\exp\langle\Phi\rangle$ is the string
coupling expressed by the dilaton vacuum expectation value
$\langle\Phi\rangle$ and $2\pi\alpha'$ is the inverse string tension.
$G_{mn}= \p_m X^M\p_n X^N g_{MN}(X)$, $B_{mn}= \p_m X^M\p_n X^N b_{MN}(X)$
and $\Phi(X)$ are the pullbacks of the background metric, antisymmetric
tensor and dilaton to the D$p$-brane worldvolume, while $F_{mn}(\xi)$ is
the field strength of the worldvolume $U(1)$ gauge field $A_m(\xi)$:
$F_{mn}=2\p_{[m}A_{n]}$. The parameter $a(p,D)$ depends on the brane and
space-time dimensions $p$ and $D$, respectively.

A D$p$-brane action, which generalizes the Polyakov type $p$-brane action,
has been introduced in \cite{AZH97}. Namely, the
action, classically equivalent to (\ref{dbia}), is given by \ba\nl S^{AZH}
&=& -\frac{T_{D_p}}{2}\int d^{p+1} \xi\Bigl\{ e^{-a(p,D)\Phi}
\sqrt{-\mathcal{K}} \left[\mathcal{K}^{mn} \left( G_{mn}+B_{mn}+
2\pi\alpha'F_{mn}\right)-(p-1)\right]\\ \nl &-&2\frac{
\varepsilon^{m_1\ldots m_{p+1}}}{(p+1)!}
\p_{m_1}X^{M_1}\ldots\p_{m_{p+1}}X^{M_{p+1}} c_{M_1\ldots
M_{p+1}}\Bigr\},\ea where $\mathcal{K}$ is the determinant of the matrix
$\mathcal{K}_{mn}$, $\mathcal{K}^{mn}$ is its inverse, and these matrices
have symmetric as well as antisymmetric part \ba\nl
\mathcal{K}^{mn}=\mathcal{K}^{(mn)}+\mathcal{K}^{[mn]},\ea where the symmetric
part $\mathcal{K}^{(mn)}$ is the analogue of the auxiliary metric
$\gamma^{mn}$ in the $p$-brane action (\ref{pa}).

Again, none of these actions satisfy all our requirements. In the same way
as in the $p$-brane case, just considered, one can prove that the action
\ba\label{oda} S_{Dp}&=&\int d^{p+1}\xi\mathcal{L}_{Dp}=\int d^{p+1}\xi
\frac{e^{-a\Phi}}{4\lambda^0}\Bigl[G_{00}-2\lambda^i G_{0i} +
\left(\lambda^i\lambda^j-\kappa^i\kappa^j\right)G_{ij}\\ \nl
&-&\left(2\lambda^0T_{D_p}\right)^2\mathbf{G} +2\kappa^i\left(
\mathcal{F}_{0i}-\lambda^j\mathcal{F}_{ji}\right)
+4\lambda^0T_{D_p}e^{a\Phi} c_{M_0\ldots M_p}\p_0
X^{M_0}\ldots\p_pX^{M_p}\Bigr],\\ \nl &&\mathcal{F}_{mn}=B_{mn} +
2\pi\alpha'F_{mn},\ea which possesses the necessary properties, is
classically equivalent to the action (\ref{dbia}). Here additional
Lagrange multipliers $\kappa^i$ are introduced, in order to linearize the
quadratic term \ba\nl
\left(\mathcal{F}_{0i}-\lambda^k\mathcal{F}_{ki}\right)
\left(G^{-1}\right)^{ij}
\left(\mathcal{F}_{0j}-\lambda^l\mathcal{F}_{lj}\right)\ea arising in the
action. For other actions of this type, see \cite{LU97} - \cite{GL98}.

Varying the action $S_{Dp}$ with respect to Lagrange multipliers $\lambda^m$,
$\kappa^i$, and requiring these variations to vanish,
one obtains the constraints
\ba\label{Dpbic0} &&G_{00}-2\lambda^{j}G_{0j}+
\left(\lambda^{i}\lambda^{j}-\kappa^i\kappa^j\right)G_{ij}
+\left(2\lambda^0T_{Dp}\right)^2\mathbf{G}
+2\kappa^i\left(\mathcal{F}_{0i}-\lambda^j\mathcal{F}_{ji}\right)=0,\\
\label{Dpbicj}
&&G_{0j}-\lambda^{i}G_{ij}=\kappa^i\mathcal{F}_{ij}\\
\label{Dpbick}
&&\mathcal{F}_{0j}-\lambda^i\mathcal{F}_{ij}=\kappa^i G_{ij}.\ea
Instead with the constraint (\ref{Dpbic0}), we will work with the simpler one
\ba\label{Dpbic0'} G_{00}-2\lambda^{j}G_{0j}+
\left(\lambda^{i}\lambda^{j}+\kappa^i\kappa^j\right)G_{ij}
+\left(2\lambda^0T_{Dp}\right)^2\mathbf{G}=0,\ea
which is obtained by inserting (\ref{Dpbick}) into (\ref{Dpbic0}).

We will use the gauge $(\lambda^m, \kappa^i)=constants$ and for simplicity,
we will restrict our considerations to constant dilaton
$\Phi=\Phi_0$ and constant electro-magnetic field $F_{mn}=F^o_{mn}$ on the
D$p$-brane worldvolume. In this case, the equations of motion for $X^M$,
following from (\ref{oda}), are
\ba\nl &&g_{LN}\Bigl[\Di\Dj X^N - \left(2\lambda^0T_{Dp}\right)^2
\p_i\left(\mathbf{G}G^{ij}\p_j X^N\right)
-\kappa^i\kappa^j\p_i\p_j X^N\Bigr]\\
\nl &&+\Gamma_{L,MN}\Bigl[\Di X^M \Dj X^N
\\ \label{Dpbem} &&- \left(2\lambda^0T_{Dp}\right)^2
\mathbf{G}G^{ij}\p_i X^M \p_j X^N-\kappa^i\kappa^j\p_i X^M\p_j X^N\Bigr]\\
\nl &&=2\la^0 T_{Dp}e^{a\Phi_0} H^{\mathbf{c}}_{LM_0\ldots
M_{p}}\p_0 X^{M_0}\ldots \p_p X^{M_p}
+H_{LMN}\kappa^j\Di X^M\p_j X^N,\ea
where $ H^{\mathbf{c}}_{p+2}=dc_{p+1}$ and $H_3=db_2$ are the corresponding
field strengths.

\setcounter{equation}{0}
\section{Exact solutions in general backgrounds}

The main idea in the mostly used approach for obtaining exact solutions of
the probe branes equations of motion in variable external fields is to
reduce the problem to a particle-like one, and even more - to solving one
dimensional dynamical problem, if possible. To achieve this, one must get
rid of the dependence on the spatial worldvolume coordinates $\xi^i$. To
this end, since the brane actions contain the first derivatives $\p_i
X^M$, the brane coordinates $X^M(\xi^m)$ have to depend on $\xi^i$ at most
linearly: \ba\label{mga} X^M(\xi^0,\xi^i)=\Lambda^M_i \xi^i +
Y^M(\xi^0),\h \Lambda^M_i  \mbox{ are arbitrary constants}.\ea Besides,
the background fields entering the action depend implicitly on $\xi^i$
through their dependence on $X^M$. If we choose $\Lambda^M_i =0$ in
(\ref{mga}), the connection with the $p$-brane setting will be lost. If we
suppose that the background fields do not depend on $X^M$, the result will
be constant background, which is not interesting in the case under
consideration. The {\it compromise} is to accept that the external fields
depend only on part of the coordinates, say $X^a$, and to set namely for
this coordinates $\Lambda^a_i =0$. In other words, we propose the ansatz
($X^M=(X^\mu,X^a)$): \ba\label{ogac} &&X^\mu(\xi^0,\xi^i)=\Lambda^\mu_j
\xi^j + Y^\mu(\xi^0),\h X^a(\xi^0,\xi^i)=Y^a(\xi^0),
\\ \label{ogab} &&\p_\mu g_{MN}=0,\h \p_\mu b_{MN}=0,\h
\p_\mu b_{M_0\ldots M_p}=0,\h \p_\mu c_{M_0\ldots M_p}=0.\ea
The resulting reduced Lagrangian density will
depend only on $\xi^0=\tau$ if the Lagrange multipliers $\lambda^m$,
$\kappa^i$ do not depend on $\xi^i$. Actually, this property follows from
their equations of motion, from where they can be expressed through
quantities depending only on the temporal worldvolume parameter $\tau$.

Thus, we have obtained the general conditions, under which the probe
branes dynamics reduces to the particle-like one. However, we will not
start our considerations relaying on the generic ansatz (\ref{ogac}).
Instead, we will begin in the framework of the commonly used in ten
space-time dimensions {\it static gauge}: $X^m(\xi^n)=\xi^m$. The latter
is a particular case of (\ref{ogac}), obtained under the following
restrictions: \ba\label{sgac} &&\mbox{(1):} \mu = i = 1,\ldots,p;\h
\mbox{(2):} \Lambda^\mu_j=\Lambda^i_j=\delta^i_j;\\ \nl &&\mbox{(3):}
Y^\mu(\tau) = Y^i(\tau) = 0;\h \mbox{(4):} Y^0(\tau)=\tau\in \{Y^a\}.\ea
Therefore, the static gauge is appropriate for backgrounds which may
depend on $X^0=Y^0(\tau)$, but must be independent on $X^i$, $(i=1,\ldots
p)$. Such properties are not satisfactory in the lower dimensions. For
instance, in four dimensional black hole backgrounds, the metric depends
on $X^1$, $X^2$ and the static gauge ansatz does not work. That is why,
our next step is to consider the probe branes dynamics in the framework of
the ansatz \ba\label{imac} X^\mu(\tau,\xi^i)=\Lambda^\mu_m \xi^m=
\Lambda^\mu_0\tau+\Lambda^\mu_i \xi^i,\h X^a(\tau,\xi^i)=Y^a(\tau),\ea
which is obtained from (\ref{ogac}) under the restriction $Y^\mu(\tau) =
\Lambda^\mu_0\tau$. Here, for the sake of symmetry between the worldvolume
coordinates $\xi^0=\tau$ and $\xi^i$, we have included in $X^\mu$ a term
linear in $\tau$. At any time, one can put $\Lambda^\mu_0=0$ and the
corresponding terms in the formulas will disappear. Further, we will refer
to the ansatz (\ref{imac}) as {\it linear gauges}, as far as $X^\mu$ are
linear combinations of $\xi^m$ with {\it arbitrary} constant coefficients.

Finally, we will investigate the classical branes dynamics by using the general
ansatz (\ref{ogac}), rewritten in the form
\ba\label{wgac} X^\mu(\tau,\xi^i)=\Lambda^\mu_0\tau+\Lambda^\mu_i \xi^i +
Y^\mu(\tau),\h X^a(\tau,\xi^i)=Y^a(\tau).\ea
Compared with (\ref{ogac}), here we have separated the linear part of $Y^\mu$
as in the previous ansatz (\ref{imac}). This will allow us to compare the role
of the term $\Lambda^\mu_0\tau$ in these two cases.

\subsection{Static gauge dynamics}
Here we begin our analysis of the probe branes dynamics in the framework
of the {\it static gauge} ansatz. In order not to introduce too many type of
indices, we will denote with $Y^a$, $Y^b$, etc., the coordinates, which are
{\it not fixed by the gauge}. However, one have not to forget that {\it by
definition}, $Y^a$ are the coordinates on which the background fields can
depend. In {\it static gauge}, according to (\ref{sgac}), one of this
coordinates, the temporal one $Y^0(\tau)$, is {\it fixed} to coincide with
$\tau$. Therefore, in this gauge, the remaining coordinates $Y^a$ are
{\it spatial} ones in space-times with signature $(-,+,\ldots,+)$.

\subsubsection{Probe $p$-branes}
In static gauge, and under the conditions (\ref{ogab}), the
action (\ref{oa}) reduces to (the over-dot is used for $d/d\tau$)
\ba\label{osga} &&S_p^{SG}=\int d\tau L_p^{SG}(\tau),\h V_p = \int d^p\xi,\\
\nl &&L_p^{SG}(\tau)= \frac{V_p}{4\lambda^0}\Bigl\{g_{ab}(Y^a)\dot{Y^a}\dot{Y^b} +
2\left[g_{0a}(Y^a) -\lambda^i g_{ia}(Y^a)+ 2\lambda^0T_p b_{a1\ldots p}(Y^a)
\right]\dot{Y^a}\\ \nl &&+ g_{00}(Y^a)-2\lambda^i g_{0i}(Y^a)
+ \lambda^i\lambda^j g_{ij}(Y^a)- \left(2\lambda^0 T_p\right)^2
\det(g_{ij}(Y^a))+ 4\lambda^0 T_p b_{01\ldots p}(Y^a)\Bigr\}.\ea
To have finite action, we require the fraction
$V_p/\lambda^0$ to be finite one. For example, in the string case ($p=1$)
and in conformal gauge ($\lambda^1=0, \left(2\lambda^0 T_1\right)^2=1$),
this means that the quantity $V_1/\alpha'= 2\pi V_1 T_1$ must be finite.

The constraints derived from the action (\ref{osga}) are: \ba\label{c0s}
&&g_{ab}\dot{Y^a}\dot{Y^b} + 2\left(g_{0a}-\lambda^i
g_{ia}\right)\dot{Y^a}+ g_{00}-2\lambda^i g_{0i} + \lambda^i\lambda^j
g_{ij}+\left(2\lambda^0T_p\right)^2 \det(g_{ij})=0,\\ \label{cis}
&&g_{ia}\dot{Y^a}+g_{i0}-g_{ij}\lambda^j=0.\ea

The Lagrangian $L_p^{SG}$ does not depend on $\tau$ explicitly, so the
energy $E_p=p^{SG}_a \dot{Y}^a-L_p^{SG}$ is conserved:
\ba\nl &&g_{ab}\dot{Y^a}\dot{Y^b}-
g_{00}+2\lambda^i g_{0i} - \lambda^i\lambda^j g_{ij}+ \left(2\lambda^0
T_p\right)^2 \det(g_{ij})- 4\lambda^0 T_p b_{01\ldots p}= \frac{4\lambda^0
E_p}{V_p}=constant.\ea
With the help of the constraints, we can replace this
equality by the following one \ba\label{p0} g_{0a}\dot{Y^a} +
g_{00}-\lambda^ig_{i0}+2\lambda^0 T_p b_{01\ldots p}=- \frac{2\lambda^0
E_p}{V_p}.\ea

To clarify the physical meaning of the equalities (\ref{cis}) and (\ref{p0}),
we compute the momenta (\ref{nm}) in static gauge
\ba\label{gms} 2\la^0 P^{SG}_M =g_{Ma}\dot{Y}^a+g_{M0}-\la^i g_{Mi}
+2\la^0T_p b_{M1\ldots p}.\ea
The comparison of (\ref{gms}) with (\ref{p0}) and (\ref{cis}) shows that
$P^{SG}_0=-E_p/V_p =const$ and $P^{SG}_i =const=0$. Inserting these conserved
momenta into (\ref{c0s}), we obtain the {\it effective} constraint
\ba\label{ecs} g_{ab}\dot{Y^a}\dot{Y^b}=\mathcal{U}^{S},\ea
where
\ba\nl \mathcal{U}^{S}=-\left(2\lambda^0 T_p\right)^2
\det(g_{ij})+ g_{00}-2\lambda^i g_{0i} + \lambda^i\lambda^j g_{ij}
+ 4\lambda^0\left(T_p b_{01\ldots p}+E_p/V_p\right).\ea

In the gauge $\lambda^m = constants$, the equations
of motion following from $S_p^{SG}$
(or from (\ref{pbem}) after imposing the static gauge) take the form:
\ba\label{ems} g_{ab}\ddot{Y^b}
+\Gamma_{a,bc}\dot{Y^b}\dot{Y^c}=\frac{1}{2}\p_a \mathcal{U}^{S} +
2\p_{[a}\mathcal{A}_{b]}^{S}\dot{Y^b},\ea
where
\ba\nl \mathcal{A}_{a}^{S}=g_{a0}-\lambda^i g_{ai}
+2\lambda^0 T_p b_{a1\ldots p}.\ea
Thus, in general, the time evolution of the reduced dynamical system does not
correspond to a geodesic motion. The deviation from the geodesic trajectory
is due to the appearance of the {\it effective} scalar potential
$\mathcal{U}^{S}$ and of the field strength $2\p_{[a}\mathcal{A}^{S}_{b]}$ of
the {\it effective} $U(1)$-gauge potential $\mathcal{A}_{a}^{S}$. In addition,
our dynamical system is subject to the {\it effective} constraint (\ref{ecs}).

\subsubsection{Probe D$p$-branes}
In static gauge, and for background fields independent of the coordinates
$X^i$ (conditions(\ref{ogab})), the reduced Lagrangian, obtained from
(\ref{oda}), is given by \ba\nl L_{Dp}^{SG}(\tau)&=& \frac{V_{Dp}
e^{-a\Phi_0}}{4\lambda^0} \Bigl[g_{ab}\dot{Y^a}\dot{Y^b}+
g_{00}-2\lambda^i g_{0i} +
\left(\lambda^i\lambda^j-\kappa^i\kappa^j\right)g_{ij}\\ \nl &+&
2\left(g_{0a} -\lambda^i g_{ia}+ 2\lambda^0T_{Dp} e^{a\Phi_0}c_{a1\ldots
p} +\kappa^i b_{ai}\right)\dot{Y^a}\\ \nl &-& \left(2\lambda^0
T_{Dp}\right)^2 \det(g_{ij})+ 4\lambda^0 T_{Dp} e^{a\Phi_0}c_{01\ldots
p}\\ \nl &+&2\kappa^i \left(b_{0i}-\lambda^j b_{ji}\right)
+4\pi\alpha'\kappa^i \left(F^o_{0i}-\lambda^j F^o_{ji}\right)\Bigr].\ea As
we already mentioned at the end of Section 2, we restrict our
considerations to the case of constant dilaton $\Phi=\Phi_0$ and constant
electro-magnetic field $F^o_{mn}$ on the D$p$-brane worldvolume.

Now, the constraints (\ref{Dpbic0'}), (\ref{Dpbicj}) and (\ref{Dpbick})
take the form
\ba\label{Dpbic0's} &&g_{ab}\dot{Y^a}\dot{Y^b} + 2\left(g_{0a}-\lambda^i
g_{ia}\right)\dot{Y^a}+ g_{00}-2\lambda^{i}g_{0i}\\ \nl&&+
\left(\lambda^{i}\lambda^{j}+\kappa^i\kappa^j\right)g_{ij}
+\left(2\lambda^0T_{Dp}\right)^2\det(g_{ij})=0,\\
\label{Dpbicjs} &&g_{ja}\dot{Y^a}+g_{0j}-\lambda^{i}g_{ij}
-\kappa^i b_{ij}=2\pi\alpha'\kappa^i F^o_{ij}\\
\nl &&b_{aj}\dot{Y^a}+b_{0j}-\lambda^{i}b_{ij}
-\kappa^i g_{ij}=-2\pi\alpha'\left(F^o_{0j}-\la^i F^o_{ij}\right).\ea

The reduced Lagrangian $L_{Dp}^{SG}$ does not depend on $\tau$ explicitly.
As a consequence, the energy $E_{Dp}$ is conserved: \ba\nl &&g_{ab}\dot{Y^a}
\dot{Y^b}-g_{00}+2\lambda^i g_{0i} -
\left(\lambda^i\lambda^j - \kappa^i\kappa^j\right)g_{ij}+
\left(2\lambda^0 T_{Dp}\right)^2 \det(g_{ij})-
4\lambda^0 T_{Dp}e^{a\Phi_0} c_{01\ldots p}
\\ \nl&&-2\kappa^i\left(b_{0i}-\lambda^j b_{ji}\right) - 4\pi\alpha'\kappa^i
\left(F^o_{0i}-\lambda^j F^o_{ji}\right)
= \frac{4\lambda^0 E_{Dp}}{V_{Dp}}e^{a\Phi_0}=constant.\ea
By using the constraints (\ref{Dpbic0's}) and (\ref{Dpbicjs}),
the above equality can be replaced by the following one
\ba\label{Dp0} g_{0a}\dot{Y^a} +
g_{00}-\lambda^ig_{i0}+2\lambda^0 T_{Dp}e^{a\Phi_0} c_{01\ldots p}
+\kappa^i\left(b_{0i}+2\pi\alpha' F^o_{0i}\right)
=- \frac{2\lambda^0 E_{Dp}}{V_{Dp}}e^{a\Phi_0}.\ea

Now, we compute the momenta, obtained from the initial action (\ref{oda}),
in static gauge
\ba\label{Dgms} 2\la^0 e^{a\Phi_0} P^{SG}_M =g_{Ma}\dot{Y}^a+g_{M0}
-\la^j g_{Mj} +2\la^0T_{Dp}e^{a\Phi_0} c_{M1\ldots p}
+\kappa^j b_{Mj}.\ea
Comparing (\ref{Dgms}) with (\ref{Dp0}) and (\ref{Dpbicjs}), one finds that
\ba\nl &&P_0^{SG}=-\left(\frac{E_{Dp}}{V_{Dp}}+\frac{\pi\alpha'}{\la^0}
e^{-a\Phi_0}\kappa^j F^o_{0j}\right)=constant,
\\ \nl &&P^{SG}_i=-\frac{\pi\alpha'}{\la^0}
e^{-a\Phi_0}\kappa^j F^o_{ij} =constants .\ea
As in the $p$-brane case, not only the energy, but also the spatial components
of the momenta $P^{SG}_i$, along the $X^i$ coordinates, are conserved.
In the D$p$-brane case however, $P^{SG}_i$ are not identically zero due the
existence of a constant worldvolume magnetic field $F^o_{ij}$.

Inserting (\ref{Dpbicjs}) and (\ref{Dp0}) into (\ref{Dpbic0's}),
one obtains the effective constraint
\ba\label{Decs} g_{ab}\dot{Y^a}\dot{Y^b} =\mathcal{U}^{DS},\ea
where
\ba\nl &&\mathcal{U}^{DS}=
-\left(2\lambda^0 T_{Dp}\right)^2 \det(g_{ij})+ g_{00}-2\lambda^i g_{0i} +
\left(\lambda^i\lambda^j-\kappa^i\kappa^j\right)g_{ij}\\ \nl &&+4\lambda^0
e^{a\Phi_0}\left(T_{Dp} c_{01\ldots p}+\frac{E_{Dp}}{V_{Dp}}\right)
+2\kappa^i\left(b_{0i}-\lambda^j b_{ji}\right)+4\pi\alpha'\kappa^i
\left(F^o_{0i}-\lambda^j F^o_{ji}\right).\ea

In the gauge $(\lambda^m,\kappa^i) = constants$, the equations
of motion following from $L_{Dp}^{SG}$
(or from (\ref{Dpbem}) after using the static gauge ansatz) take the form:
\ba\label{Dems} g_{ab}\ddot{Y^b}
+\Gamma_{a,bc}\dot{Y^b}\dot{Y^c}=\frac{1}{2}\p_a \mathcal{U}^{DS} +
2\p_{[a}\mathcal{A}_{b]}^{DS}\dot{Y^b},\ea
where
\ba\nl \mathcal{A}_{a}^{DS}=g_{a0}-\lambda^i g_{ai}
+2\lambda^0 T_{Dp} e^{a\Phi_0}c_{a1\ldots p}+\kappa^i b_{ai}.\ea

It is obvious that the equations of motion (\ref{ems}), (\ref{Dems}) and the
effective constraints (\ref{ecs}), (\ref{Decs}) have the {\it same form} for
$p$-branes and for D$p$-branes. The difference is in the explicit expressions
for the {\it effective} scalar and 1-form gauge potentials.

\subsection{Branes dynamics in linear gauges}
Now we will repeat our analysis of the probe branes dynamics in the
framework of the more general {\it linear gauges}, given by the ansatz
(\ref{imac}). The {\it static gauge} is a particular case of the {\it
linear gauges}, corresponding to the following restrictions: \ba\nl
&&\mbox{(1):} \mu = i = 1,\ldots ,p;\h \mbox{(2):}
\Lambda^\mu_0=\Lambda^i_0= 0;\\ \nl &&\mbox{(3):}
\Lambda^\mu_j=\Lambda^i_j=\delta^i_j;\h \mbox{(4):} Y^0(\tau)=\tau\in
\{Y^a\}.\ea

\subsubsection{Probe $p$-branes}
In linear gauges, and under the conditions (\ref{ogab}), one obtains the
following reduced Lagrangian, arising from the action (\ref{oa})
\ba\label{olga} &&L_p^{LG}(\tau)=
\frac{V_p}{4\lambda^0}\Bigl\{g_{ab}\dot{Y^a}\dot{Y^b} +
2\Bigl[\left(\Lambda_0^\mu-\lambda^i\Lambda_i^\mu\right)g_{\mu a}+
2\lambda^0T_p B_{a1\ldots p}\Bigr]\dot{Y^a}\\
\nl &&+\left(\Lambda_0^\mu-\lambda^i\Lambda_i^\mu\right)
\left(\Lambda_0^\nu-\lambda^j\Lambda_j^\nu\right)g_{\mu\nu}- \left(2\lambda^0
T_p\right)^2 \det(\Lambda_i^\mu\Lambda_j^\nu g_{\mu\nu})\\
\nl &&+ 4\lambda^0T_p\Lambda_0^{\mu}B_{\mu 1\ldots p}\Bigr\},
\h B_{M1\ldots p}\equiv b_{M\mu_1\ldots\mu_p}
\Lambda_1^{\mu_1}\ldots\Lambda_p^{\mu_p}.\ea

The constraints derived from the Lagrangian (\ref{olga}) are:
\ba\label{c0l}
&&g_{ab}\dot{Y^a}\dot{Y^b} +
2\left(\Lambda_0^\mu-\lambda^i\Lambda_i^\mu\right)g_{\mu a}\dot{Y^a}+
\left(\Lambda_0^\mu-\lambda^i\Lambda_i^\mu\right) \\
\nl &&\times
\left(\Lambda_0^\nu-\lambda^j\Lambda_j^\nu\right)g_{\mu\nu}+\left(2\lambda^0
T_p\right)^2 \det(\Lambda_i^\mu\Lambda_j^\nu g_{\mu\nu})=0,\\
\label{cil} &&\Lambda_i^\mu\left[g_{\mu a}\dot{Y^a}+
\left(\Lambda_0^\nu-\lambda^j\Lambda_j^\nu\right)g_{\mu\nu}\right]=0.\ea

The Lagrangian $L_p^{LG}$ does not depend on $\tau$ explicitly, so the
energy $E_p=p^{LG}_a \dot{Y}^a-L_p^{LG}$ is conserved:
\ba\nl &&g_{ab}\dot{Y^a}\dot{Y^b}-\left(\Lambda_0^\mu-
\lambda^i\Lambda_i^\mu\right)\left(\Lambda_0^\nu-\lambda^j\Lambda_j^
\nu\right)g_{\mu\nu}+ \left(2\lambda^0 T_p\right)^2
\det(\Lambda_i^\mu\Lambda_j^\nu g_{\mu\nu})\\ \nl
&&-4\lambda^0T_p\Lambda_0^{\mu}B_{\mu 1\ldots p}=
\frac{4\lambda^0 E_p}{V_p}=constant.\ea
With the help of the constraints (\ref{c0l}) and (\ref{cil}), one can replace
this equality by the following one
\ba\label{p0l} \Lambda_0^\mu\left[g_{\mu a}\dot{Y^a}+
\left(\Lambda_0^\nu-\lambda^j\Lambda_j^\nu\right)g_{\mu\nu}
+2\lambda^0T_p B_{\mu 1\ldots p}\right]=-\frac{2\lambda^0 E_p}{V_p}.\ea

In linear gauges, the momenta (\ref{nm}) take the form
\ba\label{gml} 2\la^0 P^{LG}_M =g_{Ma}\dot{Y}^a+
\left(\Lambda_0^\nu-\lambda^j\Lambda_j^\nu\right)g_{M\nu}
+2\la^0T_p B_{M1\ldots p}.\ea
The comparison of (\ref{gml}) with (\ref{p0l}) and (\ref{cil}) gives
\ba\nl \La^\mu_0 P_\mu^{LG}=-\frac{E_p}{V_p} = constant,\h
\La^\mu_i P_\mu^{LG} = constants=0.\ea
Therefore, in the linear gauges, the projections of the momenta $P^{LG}_\mu$
onto $\La^\mu_n$ are conserved. Moreover, as far as the Lagrangian
(\ref{olga}) does not depend on the coordinates $X^\mu$, the corresponding
conjugated momenta $P^{LG}_\mu$ are also conserved.

Inserting (\ref{p0l}) and (\ref{cil}) into (\ref{c0l}), we obtain the
{\it effective} constraint
\ba\nl g_{ab}\dot{Y^a}\dot{Y^b}=\mathcal{U}^{L},\ea
where the {\it effective} scalar potential is given by
\ba\nl &&\mathcal{U}^{L}=-\left(2\lambda^0 T_p\right)^2
\det(\Lambda_i^\mu\Lambda_j^\nu g_{\mu\nu})+
\left(\Lambda_0^\mu-\lambda^i\Lambda_i^\mu\right)
\left(\Lambda_0^\nu-\lambda^j\Lambda_j^\nu\right)g_{\mu\nu}\\ \nl
&&+ 4\lambda^0\left(T_p \Lambda_0^{\mu}B_{\mu 1\ldots p}
+\frac{E_p}{V_p}\right).\ea

In the gauge $\lambda^m = constants$, the equations
of motion following from $L_p^{LG}$ take the form:
\ba\nl g_{ab}\ddot{Y^b}
+\Gamma_{a,bc}\dot{Y^b}\dot{Y^c}=\frac{1}{2}\p_a \mathcal{U}^{L} +
2\p_{[a}\mathcal{A}_{b]}^{L}\dot{Y^b},\ea
where
\ba\nl \mathcal{A}_{a}^{L}=
\left(\Lambda_0^\mu-\lambda^i\Lambda_i^\mu\right)g_{a\mu}+
2\lambda^0T_p B_{a1\ldots p},\ea
is the {\it effective} 1-form gauge potential, generated by the non-diagonal
components $g_{a\mu}$ of the background metric and by the components
$b_{a\mu_1 \ldots \mu_p}$ of the background $(p+1)$-form gauge field.

\subsubsection{Probe D$p$-branes}
In linear gauges, and for background fields independent of the coordinates
$X^\mu$ (conditions(\ref{ogab})), the reduced Lagrangian, obtained from
(\ref{oda}), is given by \ba\nl L_{Dp}^{LG}(\tau)&=&
\frac{V_{Dp} e^{-a\Phi_0}}{4\lambda^0} \Bigl\{g_{ab}\dot{Y^a}\dot{Y^b}+
\left[\left(\Lambda_0^\mu-\lambda^i\Lambda_i^\mu\right)
\left(\Lambda_0^\nu-\lambda^j\Lambda_j^\nu\right)
-\kappa^i\kappa^j\Lambda_i^\mu\Lambda_j^\nu\right]g_{\mu\nu}\\ \nl &+&
2\left[\left(\Lambda_0^\mu-\lambda^i\Lambda_i^\mu\right)g_{\mu a}
+ 2\lambda^0T_{Dp} e^{a\Phi_0}C_{a1\ldots p}
+\kappa^i\La^\mu_i b_{a\mu}\right]\dot{Y^a}\\
\nl &-& \left(2\lambda^0 T_{Dp}\right)^2
\det(\Lambda_i^\mu\Lambda_j^\nu g_{\mu\nu})+ 4\lambda^0 T_{Dp} e^{a\Phi_0}
\La^\mu_0 C_{\mu 1\ldots p}\\ \nl &-&2\kappa^i\La^\mu_i
\left(\La^\nu_0-\lambda^j \La^\nu_j\right)b_{\mu\nu} +4\pi\alpha'\kappa^i
\left(F^o_{0i}-\lambda^j F^o_{ji}\right)\Bigr\},\ea
where the following shorthand notation has been introduced
\ba\nl C_{M1\ldots p}\equiv
c_{M\mu_1\ldots\mu_p}\La^{\mu_1}_1\ldots\La^{\mu_p}_p.\ea

Now, the constraints (\ref{Dpbic0'}), (\ref{Dpbicj}), and (\ref{Dpbick})
take the form \ba\label{Dpbic0'l} &&g_{ab}\dot{Y^a}\dot{Y^b} +
2\left(\Lambda_0^\mu-\lambda^i\Lambda_i^\mu\right)g_{\mu a}\dot{Y^a}
+\left(2\lambda^0T_{Dp}\right)^2 \det(\Lambda_i^\mu\Lambda_j^\nu
g_{\mu\nu})\\ \nl
&&+\left[\left(\Lambda_0^\mu-\lambda^i\Lambda_i^\mu\right)
\left(\Lambda_0^\nu-\lambda^j\Lambda_j^\nu\right)
+\kappa^i\kappa^j\Lambda_i^\mu\Lambda_j^\nu\right]g_{\mu\nu}=0,
\\ \label{Dpbicjl} &&\Lambda_i^\mu\left[g_{\mu a}\dot{Y^a}+
\left(\Lambda_0^\nu-\lambda^j\Lambda_j^\nu\right)g_{\mu\nu}+
\kappa^j\La^\nu_j b_{\mu\nu}\right]=2\pi\alpha'\kappa^j F^o_{ji}\\
\nl &&\Lambda_i^\mu\left[b_{\mu a}\dot{Y^a}+
\left(\Lambda_0^\nu-\lambda^j\Lambda_j^\nu\right)b_{\mu\nu}+\kappa^j\La^\nu_j
g_{\mu\nu}\right]=2\pi\alpha'\left(F^o_{0i}-\la^j F^o_{ji}\right).\ea

The reduced Lagrangian $L_{Dp}^{LG}$ does not depend on $\tau$ explicitly.
As a consequence, the energy $E_{Dp}$ is conserved: \ba\nl &&g_{ab}\dot{Y^a}
\dot{Y^b}-\left[\left(\Lambda_0^\mu-\lambda^i\Lambda_i^\mu\right)
\left(\Lambda_0^\nu-\lambda^j\Lambda_j^\nu\right)
-\kappa^i\kappa^j\Lambda_i^\mu\Lambda_j^\nu\right]g_{\mu\nu}\\ \nl
&&+\left(2\lambda^0 T_{Dp}\right)^2 \det(\Lambda_i^\mu\Lambda_j^\nu
g_{\mu\nu})- 4\lambda^0 T_{Dp}e^{a\Phi_0}\La^\mu_0 C_{\mu 1\ldots p}
\\ \nl&&-2\kappa^i\La^\mu_i
\left(\La^\nu_0-\lambda^j \La^\nu_j\right)b_{\mu\nu} - 4\pi\alpha'\kappa^i
\left(F^o_{0i}-\lambda^j F^o_{ji}\right)
= \frac{4\lambda^0 E_{Dp}}{V_{Dp}}e^{a\Phi_0}=constant.\ea
By using the constraints (\ref{Dpbic0'l}) and (\ref{Dpbicjl}),
the above equality can be replaced by the following one
\ba\label{Dp0l} &&\Lambda_0^\mu\left[g_{\mu a}\dot{Y^a}+
\left(\Lambda_0^\nu-\lambda^j\Lambda_j^\nu\right)g_{\mu\nu}
+2\lambda^0T_{Dp}e^{a\Phi_0} C_{\mu 1\ldots p}
+\kappa^j\La^\nu_j b_{\mu\nu}\right]\\ \nl
&&+2\pi\alpha'\kappa^i F^o_{0i}
=- \frac{2\lambda^0 E_{Dp}}{V_{Dp}}e^{a\Phi_0}.\ea

In linear gauges, the momenta obtained from the initial action (\ref{oda}),
are
\ba\label{Dgml} 2\la^0 e^{a\Phi_0} P^{LG}_M =
g_{Ma}\dot{Y^a}+
\left(\Lambda_0^\nu-\lambda^j\Lambda_j^\nu\right)g_{M\nu}
+2\lambda^0T_{Dp}e^{a\Phi_0} C_{M1\ldots p}
+\kappa^j\La^\nu_j b_{M\nu}.\ea
Comparing (\ref{Dgml}) with (\ref{Dp0l}) and (\ref{Dpbicjl}), one finds that
the following equalities hold
\ba\nl &&\La^\mu_0 P_\mu^{LG}=-\left(\frac{E_{Dp}}{V_{Dp}}
+\frac{\pi\alpha'}{\la^0}
e^{-a\Phi_0}\kappa^j F^o_{0j}\right)=constant,
\\ \nl &&\La^\mu_i P_\mu^{LG}=-\frac{\pi\alpha'}{\la^0}
e^{-a\Phi_0}\kappa^j F^o_{ij} =constants .\ea
They may be viewed as restrictions on the number of the
arbitrary parameters, presented in the theory.

As in the $p$-brane case, the momenta $P_\mu^{LG}$ are conserved quantities,
due to the independence of the Lagrangian on the coordinates $X^\mu$.

Inserting (\ref{Dpbicjl}) and (\ref{Dp0l}) into (\ref{Dpbic0'l}),
one obtains the effective constraint
\ba\nl g_{ab}\dot{Y^a}\dot{Y^b} =\mathcal{U}^{DL},\ea
where
\ba\nl &&\mathcal{U}^{DL}=
\left[\left(\Lambda_0^\mu-\lambda^i\Lambda_i^\mu\right)
\left(\Lambda_0^\nu-\lambda^j\Lambda_j^\nu\right)
-\kappa^i\kappa^j\Lambda_i^\mu\Lambda_j^\nu\right]g_{\mu\nu}
\\ \nl &&-\left(2\lambda^0 T_{Dp}\right)^2 \det(\Lambda_i^\mu\Lambda_j^\nu
g_{\mu\nu})+4\lambda^0
e^{a\Phi_0}\left(T_{Dp}\La^\mu_0 C_{\mu 1\ldots p}+
\frac{E_{Dp}}{V_{Dp}}\right)\\ \nl
&&+2\kappa^i\La^\mu_i\left(\La^\nu_0-\lambda^j \La^\nu_j\right)b_{\mu\nu}
+4\pi\alpha'\kappa^i\left(F^o_{0i}-\lambda^j F^o_{ji}\right).\ea

In the gauge $(\lambda^m,\kappa^i) = constants$, the equations
of motion following from $L_{Dp}^{LG}$ take the form:
\ba\nl g_{ab}\ddot{Y^b}
+\Gamma_{a,bc}\dot{Y^b}\dot{Y^c}=\frac{1}{2}\p_a \mathcal{U}^{DL} +
2\p_{[a}\mathcal{A}_{b]}^{DL}\dot{Y^b},\ea
where
\ba\nl \mathcal{A}_{a}^{DL}=
\left(\Lambda_0^\mu-\lambda^i\Lambda_i^\mu\right)g_{a\mu}
+ 2\lambda^0T_{Dp} e^{a\Phi_0}C_{a1\ldots p}
+\kappa^i\La^\mu_i b_{a\mu}.\ea

It is clear that the equations of motion and the effective constraints
have the {\it same form} for $p$-branes and for D$p$-branes in linear gauges,
as well as in static gauge. The only difference is in the explicit expressions
for the {\it effective} scalar and 1-form gauge potentials.

\subsection{Branes dynamics in the whole space-time}
Working in {\it static gauge} $X^m(\xi^n)=\xi^m$, we actually imply that
the probe branes have no dynamics along the background coordinates $x^m$. The
(proper) time evolution is possible only in the transverse directions,
described by the coordinates $x^a$.

Using the {\it linear gauges}, we have the possibility to place the probe
branes in general position with respect to the coordinates $x^\mu$,
on which the background fields do not depend. However, the real dynamics is
again in the transverse directions only.

Actually, in the framework of our approach, the probe branes can have
'full' dynamical freedom only when the ansatz (\ref{wgac}) is used,
because only then {\it all} of the brane coordinates $X^M$ are allowed to
vary {\it nonlinearly} with the proper time $\tau$. Therefore, with the
help of (\ref{wgac}), we can probe the {\it whole} space-time.

We will use the superscript $A$ to denote that the corresponding quantity is
taken on the ansatz (\ref{wgac}). It is understood that the conditions
(\ref{ogab}) are also fulfilled.

\subsubsection{Probe $p$-branes}
Now, the reduced Lagrangian obtained from the action (\ref{oa}) is given by
\ba\nl &&L_p^{A}(\tau)=
\frac{V_p}{4\lambda^0}\Bigl\{g_{MN}\dot{Y^M}\dot{Y^N} +
2\Bigl[\left(\Lambda_0^\mu-\lambda^i\Lambda_i^\mu\right)g_{\mu N}+
2\lambda^0T_p B_{N1\ldots p}\Bigr]\dot{Y^N}\\
\nl &&+\left(\Lambda_0^\mu-\lambda^i\Lambda_i^\mu\right)
\left(\Lambda_0^\nu-\lambda^j\Lambda_j^\nu\right)g_{\mu\nu}- \left(2\lambda^0
T_p\right)^2 \det(\Lambda_i^\mu\Lambda_j^\nu g_{\mu\nu})\\
\nl &&+ 4\lambda^0T_p\Lambda_0^{\mu}B_{\mu 1\ldots p}\Bigr\}.\ea

The constraints, derived from the above Lagrangian, are:
\ba\label{c0g}
&&g_{MN}\dot{Y^M}\dot{Y^N} +
2\left(\Lambda_0^\mu-\lambda^i\Lambda_i^\mu\right)g_{\mu N}\dot{Y^N}+
\left(\Lambda_0^\mu-\lambda^i\Lambda_i^\mu\right) \\
\nl &&\times
\left(\Lambda_0^\nu-\lambda^j\Lambda_j^\nu\right)g_{\mu\nu}+\left(2\lambda^0
T_p\right)^2 \det(\Lambda_i^\mu\Lambda_j^\nu g_{\mu\nu})=0,\\
\label{cig} &&\Lambda_i^\mu\left[g_{\mu N}\dot{Y^N}+
\left(\Lambda_0^\nu-\lambda^j\Lambda_j^\nu\right)g_{\mu\nu}\right]=0.\ea
The corresponding momenta are ($P_M = P^{A}_M/V_p$)
\ba\nl 2\la^0 P_M =g_{MN}\dot{Y}^N+
\left(\Lambda_0^\nu-\lambda^j\Lambda_j^\nu\right)g_{M\nu}
+2\la^0T_p B_{M1\ldots p},\ea
and part of them, $P_\mu$, are conserved
\ba\label{gmgc} g_{\mu N}\dot{Y}^N+
\left(\Lambda_0^\nu-\lambda^j\Lambda_j^\nu\right)g_{\mu\nu}
+2\la^0T_p B_{\mu 1\ldots p}= 2\la^0 P_\mu = constants ,\ea
because $L_p^{A}$ does not depend on $X^\mu$.
From (\ref{cig}) and (\ref{gmgc}), the compatibility conditions follow
\ba\label{ccs} \La^\mu_i P_\mu =0.\ea
We will regard on (\ref{ccs}) as a solution of the constraints (\ref{cig}),
which restricts the number of the arbitrary parameters
$\La^\mu_i$ and $P_\mu$. That is why from now on, we will deal only with the
constraint (\ref{c0g}).

In the gauge $\lambda^m = constants$, the equations
of motion for $Y^N$, following from $L_p^{A}$, have the form
\ba\label{emg} g_{LN}\ddot{Y^N}
+\Gamma_{L,MN}\dot{Y^M}\dot{Y^N}=\frac{1}{2}\p_L \mathcal{U}^{in} +
2\p_{[L}\mathcal{A}_{N]}^{in}\dot{Y^N},\ea
where
\ba\nl &&\mathcal{U}^{in}=-\left(2\lambda^0 T_p\right)^2
\det(\Lambda_i^\mu\Lambda_j^\nu g_{\mu\nu})+
\left(\Lambda_0^\mu-\lambda^i\Lambda_i^\mu\right)
\left(\Lambda_0^\nu-\lambda^j\Lambda_j^\nu\right)g_{\mu\nu}\\ \nl
&&+ 4\lambda^0T_p \Lambda_0^{\mu}B_{\mu 1\ldots p} ,\\
\nl &&\mathcal{A}_{N}^{in}=
\left(\Lambda_0^\mu-\lambda^i\Lambda_i^\mu\right)g_{N\mu}+
2\lambda^0T_p B_{N1\ldots p}.\ea
Let us first consider this part of the equations of motion (\ref{emg}), which
corresponds to $L=\la$. It follows from (\ref{ogab}) that the connection
coefficients $\Gamma_{\la,MN}$, involved in these equations, are
\ba\nl \Gamma_{\lambda,ab}=\frac{1}{2}\left(\p_ag_{b\lambda}
+\p_bg_{a\lambda}\right), \h \Gamma_{\lambda,\mu
a}=\frac{1}{2}\p_ag_{\mu\lambda},\h \Gamma_{\lambda,\mu\nu}=0.\ea
Inserting these expressions in the part of the differential equations
(\ref{emg}) corresponding to $L=\lambda$ and using that
$\dot{g}_{MN}=\dot{Y}^a\p_ag_{MN}$,
$\dot{B}_{M1\ldots p}=\dot{Y}^a\p_a B_{M1\ldots p}$,
one receives
\ba\nl \frac{d}{d\tau}\left[g_{\mu N}\dot{Y}^N+
\left(\Lambda_0^\nu-\lambda^j\Lambda_j^\nu\right)g_{\mu\nu}
+2\la^0T_p B_{\mu 1\ldots p}\right]=0.\ea
These equalities express the fact that the momenta $P_\mu$ are conserved
(compare with (\ref{gmgc})). Therefore, we have to deal only with the other
part of the equations of motion, corresponding to $L=a$
\ba\label{emga} g_{aN}\ddot{Y^N}
+\Gamma_{a,MN}\dot{Y^M}\dot{Y^N}=\frac{1}{2}\p_a \mathcal{U}^{in} +
2\p_{[a}\mathcal{A}_{N]}^{in}\dot{Y^N}.\ea

Our next task is to {\it separate the variables} $\dot{Y^\mu}$ and
$\dot{Y^a}$ in these equations and in the constraint (\ref{c0g}). To this
end, we will use the conservation laws (\ref{gmgc}) to express
$\dot{Y^\mu}$ through $\dot{Y^a}$. The result is \ba\label{ymu}
\dot{Y^\mu}=\left(g^{-1}\right)^{\mu\nu} \left[2\la^0(P_\nu - T_p B_{\nu
1\ldots p}) -g_{\nu a}\dot{Y^a}\right] -(\La^\mu_0 - \la^i\La^\mu_i).\ea
We will need also the explicit expressions for the connection coefficients
$\Gamma_{a,\mu b}$ and $\Gamma_{a,\mu\nu}$, which under the conditions
(\ref{ogab}) reduce to \ba\label{eeg} \Gamma_{a,\mu
b}=-\frac{1}{2}\left(\p_ag_{b\mu}-\p_bg_{a\mu}\right)
=-\p_{[a}g_{b]\mu},\h \Gamma_{a,\mu\nu}=-\frac{1}{2}\p_ag_{\mu\nu}.\ea By
using (\ref{ymu}) and (\ref{eeg}), after some calculations, one rewrites
the equations of motion (\ref{emga}) and the constraint (\ref{c0g}) in the
form \ba\label{mgemf} &&h_{ab}\ddot{Y}^b +
\Gamma^{\bf{h}}_{a,bc}\dot{Y}^b\dot{Y}^c = \frac{1}{2}\p_a \mathcal{U}^{A}
+ 2\p_{[a}\mathcal{A}^{A}_{b]}\dot{Y}^b,\\ \label{mgecf}
&&h_{ab}\dot{Y}^a\dot{Y}^b = \mathcal{U}^{A},\ea where a new, {\it
effective metric} appeared \ba\nl h_{ab} = g_{ab} -
g_{a\mu}(g^{-1})^{\mu\nu}g_{\nu b}.\ea $\Gamma^{\bf{h}}_{a,bc}$ is the
connection compatible with this metric \ba\nl
\Gamma^{\bf{h}}_{a,bc}=\frac{1}{2}\left(\p_b h_{ca} +\p_c h_{ba}-\p_a
h_{bc}\right).\ea The new, {\it effective} scalar and gauge potentials are
given by \ba\nl &&\mathcal{U}^{A}=-\left(2\lambda^0 T_p\right)^2
\det(\Lambda_i^\mu\Lambda_j^\nu g_{\mu\nu}) - (2\la^0)^2 \left(P_\mu-T_p
B_{\mu 1\ldots p}\right)\left(g^{-1}\right)^{\mu\nu} \left(P_\nu-T_p
B_{\nu 1\ldots p}\right),\\ \nl &&\mathcal{A}_{a}^{A}=
2\lambda^0\left[g_{a\mu}\left(g^{-1}\right)^{\mu\nu} \left(P_\nu-T_p
B_{\nu 1\ldots p}\right)+ T_p B_{a1\ldots p}\right].\ea We note that Eqs.
(\ref{mgemf}), (\ref{mgecf}), and therefore their solutions, do not depend
on the parameters $\La^\mu_0$ and $\la^i$ in contrast to the previously
considered cases. However, they have the same form as before.

\subsubsection{Probe D$p$-branes}

The reduced Lagrangian, obtained from (\ref{oda}), is given by
\ba\nl  L_{Dp}^{A}(\tau)&=&
\frac{V_{Dp} e^{-a\Phi_0}}{4\lambda^0} \Bigl\{g_{MN}\dot{Y^M}\dot{Y^N}+
\left[\left(\Lambda_0^\mu-\lambda^i\Lambda_i^\mu\right)
\left(\Lambda_0^\nu-\lambda^j\Lambda_j^\nu\right)
-\kappa^i\kappa^j\Lambda_i^\mu\Lambda_j^\nu\right]g_{\mu\nu}\\ \nl &+&
2\left[\left(\Lambda_0^\mu-\lambda^i\Lambda_i^\mu\right)g_{\mu N}
+ 2\lambda^0T_{Dp} e^{a\Phi_0}C_{N1\ldots p}
+\kappa^i\La^\mu_i b_{N\mu}\right]\dot{Y^N}\\
\nl &-& \left(2\lambda^0 T_{Dp}\right)^2
\det(\Lambda_i^\mu\Lambda_j^\nu g_{\mu\nu})+ 4\lambda^0 T_{Dp} e^{a\Phi_0}
\La^\mu_0 C_{\mu 1\ldots p}\\ \nl &-&2\kappa^i\La^\mu_i
\left(\La^\nu_0-\lambda^j \La^\nu_j\right)b_{\mu\nu} +4\pi\alpha'\kappa^i
\left(F^o_{0i}-\lambda^j F^o_{ji}\right)\Bigr\},\ea

The constraints (\ref{Dpbic0'}), (\ref{Dpbicj}), and (\ref{Dpbick}) take
the form \ba\label{Dpbic0'g} &&g_{MN}\dot{Y^M}\dot{Y^N} +
2\left(\Lambda_0^\mu-\lambda^i\Lambda_i^\mu\right)g_{\mu N}\dot{Y^N}
+\left(2\lambda^0T_{Dp}\right)^2 \det(\Lambda_i^\mu\Lambda_j^\nu
g_{\mu\nu})\\ \nl
&&+\left[\left(\Lambda_0^\mu-\lambda^i\Lambda_i^\mu\right)
\left(\Lambda_0^\nu-\lambda^j\Lambda_j^\nu\right)
+\kappa^i\kappa^j\Lambda_i^\mu\Lambda_j^\nu\right]g_{\mu\nu}=0,
\\ \label{Dpbicjg} &&\Lambda_i^\mu\left[g_{\mu N}\dot{Y^N}+
\left(\Lambda_0^\nu-\lambda^j\Lambda_j^\nu\right)g_{\mu\nu}+
\kappa^j\La^\nu_j b_{\mu\nu}\right]=2\pi\alpha'\kappa^j F^o_{ji}\\
\label{Dpbickg} &&\Lambda_i^\mu\left[b_{\mu N}\dot{Y^N}+
\left(\Lambda_0^\nu-\lambda^j\Lambda_j^\nu\right)b_{\mu\nu}+\kappa^j\La^\nu_j
g_{\mu\nu}\right]=2\pi\alpha'\left(F^o_{0i}-\la^j F^o_{ji}\right).\ea

Because of the independence of $L_{Dp}^{A}$ on $X^\mu$, the momenta
$P^D_\mu=P_\mu^{DA}/V_{Dp}$
are conserved
\ba\label{gmgcd} 2\la^0 e^{a\Phi_0} P^D_\mu =
g_{\mu N}\dot{Y^N}+
\left(\Lambda_0^\nu-\lambda^j\Lambda_j^\nu\right)g_{\mu\nu}
+2\lambda^0T_{Dp}e^{a\Phi_0} C_{\mu 1\ldots p}
+\kappa^j\La^\nu_j b_{\mu\nu} = constants.\ea
From (\ref{Dpbicjg}) and (\ref{gmgcd}), one obtains the following
compatibility conditions
\ba\nl \La^\mu_j P^D_\mu = \frac{\pi\alpha'}{\la^0}
e^{-a\Phi_0}\kappa^i F^o_{ij},\ea
which we interpret as a solution of the constraints (\ref{Dpbicjg}).

In the gauge $(\lambda^m,\kappa^i) = constants$, the equations
of motion for $Y^N$, following from $L_{Dp}^{A}$, take the form
\ba\label{emgd} g_{LN}\ddot{Y^N}
+\Gamma_{L,MN}\dot{Y^M}\dot{Y^N}=\frac{1}{2}\p_L \mathcal{U}^{Din} +
2\p_{[L}\mathcal{A}_{N]}^{Din}\dot{Y^N},\ea
where
\ba\nl &&\mathcal{U}^{Din}=-\left(2\lambda^0 T_p\right)^2
\det(\Lambda_i^\mu\Lambda_j^\nu g_{\mu\nu})+
\left[\left(\Lambda_0^\mu-\lambda^i\Lambda_i^\mu\right)
\left(\Lambda_0^\nu-\lambda^j\Lambda_j^\nu\right)
-\kappa^i\kappa^j\Lambda_i^\mu\Lambda_j^\nu\right]g_{\mu\nu}
\\ \nl &&+ 4\lambda^0 T_{Dp} e^{a\Phi_0}
\La^\mu_0 C_{\mu 1\ldots p}-2\kappa^i\La^\mu_i
\left(\La^\nu_0-\lambda^j \La^\nu_j\right)b_{\mu\nu},\\
\nl &&\mathcal{A}_{N}^{Din}=
\left(\Lambda_0^\nu-\lambda^j\Lambda_j^\nu\right)g_{N\nu}
+ 2\lambda^0T_{Dp} e^{a\Phi_0}C_{N1\ldots p}
+\kappa^j\La^\nu_j b_{N\nu}.\ea
As in the $p$-brane case, this part of the equations of motion (\ref{emgd}),
which corresponds to $L=\la$, expresses the conservation of the momenta
$P^D_\mu$, in accordance with (\ref{gmgcd}). The remaining equations of
motion, which we have to deal with, are
\ba\label{emgad} g_{aN}\ddot{Y^N}
+\Gamma_{a,MN}\dot{Y^M}\dot{Y^N}=\frac{1}{2}\p_a \mathcal{U}^{Din} +
2\p_{[a}\mathcal{A}_{N]}^{Din}\dot{Y^N}.\ea

To exclude the dependence on $\dot{Y}^\mu$ in the Eqs. (\ref{emgad}) and
in the constraints (\ref{Dpbic0'g}), (\ref{Dpbickg}), we use the
conservation laws (\ref{gmgcd}) to express $\dot{Y^\mu}$ through
$\dot{Y^a}$: \ba\label{ymud} \dot{Y^\mu}=\left(g^{-1}\right)^{\mu\nu}
\left[2\la^0 e^{a\Phi_0}(P^D_\nu - T_{Dp}C_{\nu 1\ldots p}) -g_{\nu
a}\dot{Y^a} -\kappa^j\La^\rho_j b_{\nu\rho}\right] -(\La^\mu_0 -
\la^i\La^\mu_i).\ea

By using (\ref{ymud}) and (\ref{eeg}), one can rewrite the equations of motion
(\ref{emgad}) and the constraint (\ref{Dpbic0'g}) as
\ba\label{mgemfd} &&h_{ab}\ddot{Y}^b +
\Gamma^{\bf{h}}_{a,bc}\dot{Y}^b\dot{Y}^c
= \frac{1}{2}\p_a \mathcal{U}^{DA}
+ 2\p_{[a}\mathcal{A}^{DA}_{b]}\dot{Y}^b,\\
\label{mgecfd} &&h_{ab}\dot{Y}^a\dot{Y}^b = \mathcal{U}^{DA}.\ea
Now, the {\it effective} scalar and 1-form gauge potentials are given by
\ba\nl \mathcal{U}^{DA}&=&-\left(2\lambda^0 T_p\right)^2
\det(\Lambda_i^\mu\Lambda_j^\nu g_{\mu\nu})
-\kappa^i\kappa^j\Lambda_i^\mu\Lambda_j^\nu g_{\mu\nu}
\\ \nl &-&\left[2\la^0 e^{a\Phi_0}(P^D_\mu
- T_{Dp}C_{\mu 1\ldots p})-\kappa^i\La^\la_i b_{\mu\la}\right]
\left(g^{-1}\right)^{\mu\nu}
\\ \nl&\times&\left[2\la^0 e^{a\Phi_0}(P^D_\nu
- T_{Dp}C_{\nu 1\ldots p})-\kappa^j\La^\rho_j b_{\nu\rho}\right],\\ \nl
\mathcal{A}_{a}^{DA}&=&
g_{a\mu}\left(g^{-1}\right)^{\mu\nu}\left[2\la^0 e^{a\Phi_0}(P^D_\nu
- T_{Dp}C_{\nu 1\ldots p})-\kappa^j\La^\rho_j b_{\nu\rho}\right]
\\ \nl &+& 2\lambda^0T_{Dp} e^{a\Phi_0}C_{a1\ldots p}
+\kappa^i\La^\mu_i b_{a\mu}.\ea Eqs. (\ref{mgemfd}), (\ref{mgecfd}), have
the same form as in static and linear gauges, but now they do not depend
on the parameters $\La^\mu_0$ and $\la^i$. Another difference is the
appearance of a new, {\it effective} background metric $h_{ab}$ and the
corresponding connection $\Gamma^{\bf{h}}_{a,bc}$.

In the D-brane case, we have another set of constraints (\ref{Dpbickg}),
generated by the Lagrange multipliers $\kappa^i$. With the help of
(\ref{ymud}), they acquire the form
\ba\nl &&\Bigl\{\left[b_{a\nu}-g_{a\rho}\left(g^{-1}\right)^{\rho\mu}b_{\mu\nu}
\right]\dot{Y}^a +b_{\mu\nu}\left(g^{-1}\right)^{\mu\rho}
\left[2\la^0 e^{a\Phi_0}(P^D_\rho - T_{Dp}C_{\rho 1\ldots p})-
\kappa^j\La^\la_j b_{\rho\la}\right]\\ \nl
&& -\kappa^i\La^\mu_i g_{\mu\nu}\Bigr\}
\La^\nu_j =-2\pi\alpha'\left(F^o_{0j}-\la^i F^o_{ij}\right).\ea

\subsection{Explicit solutions of the equations of motion}

All cases considered so far, have one common feature. The dynamics of the
corresponding reduced particle-like system is described by {\it effective}
equations of motion and one {\it effective} constraint, which have the
{\it same form}, independently of the ansatz used to reduce the $p$-branes
or D$p$-branes dynamics. Our aim here is to find {\it explicit exact
solutions} to them. \footnote{The additional restrictions on the
solutions, depending on the ansatz and on the type of the branes, will be
discussed in the next section.} To be able to describe all cases
simultaneously, let us first introduce some general notations.

We will search for solutions of the following system of {\it nonlinear}
differential equations \ba\label{ee} &&\mathcal{G}_{ab}\ddot{Y}^b +
\Gamma^{\mathcal{G}}_{a,bc}\dot{Y}^b\dot{Y}^c = \frac{1}{2}\p_a
\mathcal{U} + 2\p_{[a}\mathcal{A}_{b]}\dot{Y}^b,\\ \label{ec}
&&\mathcal{G}_{ab}\dot{Y}^a\dot{Y}^b = \mathcal{U},\ea where
$\mathcal{G}_{ab}$, $\Gamma^{\mathcal{G}}_{a,bc}$, $\mathcal{U}$, and
$\mathcal{A}_a$ can be as follows \ba\nl
&&\mathcal{G}_{ab}=\left(g_{ab},h_{ab}\right),\h
\Gamma^{\mathcal{G}}_{a,bc}=\left(\Gamma_{a,bc},\Gamma^{\bf{h}}_{a,bc}\right),
\\ \nl
&&\mathcal{U}=\left(\mathcal{U}^{S},\mathcal{U}^{DS},\mathcal{U}^{L},
\mathcal{U}^{DL},\mathcal{U}^{A},\mathcal{U}^{DA}\right),
\\ \nl
&&\mathcal{A}_a=\left(\mathcal{A}_{a}^{S},\mathcal{A}_{a}^{DS},
\mathcal{A}_{a}^{L},\mathcal{A}_{a}^{DL},\mathcal{A}_{a}^{A},
\mathcal{A}_{a}^{DA}\right),\ea
depending on the ansatz and on the type of the brane ($p$-brane or D$p$-brane).

Let us start with the simplest case, when the background fields depend on
only one coordinate $X^a=Y^a(\tau)$. \footnote{An example of such
background is the generalized Kasner type metric, arising in the
superstring cosmology \cite{LWC99} (see also \cite{CKR02},
\cite{EGJK02}).} In this case the Eqs. (\ref{ee}), (\ref{ec}) simplify to
($d_a\equiv d/dY^a$) \ba\label{ee1}
&&\frac{d}{d\tau}\left(\mathcal{G}_{aa}\dot{Y}^a\right)
-\frac{1}{2}d_a\mathcal{G}_{aa}\left(\dot{Y}^a\right)^2 = \frac{1}{2}d_a
\mathcal{U} ,\\ \label{ec1} &&\mathcal{G}_{aa}\left(\dot{Y}^a\right)^2 =
\mathcal{U},\ea where we have used that \ba\nl \mathcal{G}_{ab}\ddot{Y}^b
+ \Gamma^{\mathcal{G}}_{a,bc}\dot{Y}^b\dot{Y}^c
=\frac{d}{d\tau}\left(\mathcal{G}_{ab}\dot{Y}^b\right)
-\frac{1}{2}\p_a\mathcal{G}_{bc}\dot{Y}^b\dot{Y}^c.\ea After multiplying
with $2\mathcal{G}_{aa}\dot{Y}^a$ and after using the constraint
(\ref{ec1}),  the Eq. (\ref{ee1}) reduces to \ba\label{fi1}
\frac{d}{d\tau}\left[\left(\mathcal{G}_{aa}\dot{Y}^a\right)^2 -
\mathcal{G}_{aa}\mathcal{U}\right]=0.\ea The solution of (\ref{fi1}),
compatible with (\ref{ec1}), is just the constraint (\ref{ec1}). In other
words, (\ref{ec1}) is first integral of the equation of motion for the
coordinate $Y^a$. By integrating (\ref{ec1}), one obtains the following
exact probe branes solution \ba\label{tsol1}\tau\left(X^a\right)=\tau_0
\pm \int_{X_0^a}^{X^a}
\left(\frac{\mathcal{U}}{\mathcal{G}_{aa}}\right)^{-1/2}dx,\ea where
$\tau_0$ and $X_0^a$ are arbitrary constants.

When one works in the framework of the general ansatz (\ref{wgac}),
one has to also write down the solution for the remaining coordinates $X^\mu$.
It can be obtained as follows. One represents $\dot{Y}^\mu$ as
\ba\nl \dot{Y}^\mu =\frac{dY^\mu}{dY^a}\dot{Y}^a,\ea
and use this and (\ref{ec1}) in (\ref{ymu}) for the $p$-brane,
and in (\ref{ymud}) for the D$p$-brane. The result is a system of ordinary
differential equations of first order with separated variables, which
integration is straightforward. Replacing the obtained solution for
$Y^\mu(X^a)$ in the ansatz (\ref{wgac}), one finally arrives at
\ba\label{xmus1} &&X^\mu(X^a,\xi^i)=X^\mu_0 +
\La^\mu_i\left[\la^i\tau(X^a)+\xi^i\right]
\\ \nl &&-\int_{X^a_0}^{X^a}\left(g^{-1}\right)^{\mu\nu}\left[g_{\nu a}
\mp 2\la^0(P_\nu-T_p B_{\nu 1\ldots p})
\left(\frac{\mathcal{U}^A}{h_{aa}}\right)^{-1/2}\right]dx\ea
for the $p$-brane case, and at
\ba\label{xmusd1} &&X^\mu(X^a,\xi^i)=X^\mu_0 +
\La^\mu_i\left[\la^i\tau(X^a)+\xi^i\right]
\\ \nl &&-\int_{X^a_0}^{X^a}\left(g^{-1}\right)^{\mu\nu}\left\{g_{\nu a}
\mp \left[2\la^0 e^{a\Phi_0}(P^D_\nu
- T_{Dp}C_{\nu 1\ldots p})-\kappa^j\La^\rho_j b_{\nu\rho}\right]
\left(\frac{\mathcal{U}^{DA}}{h_{aa}}\right)^{-1/2}\right\}dx\ea
for the D$p$-brane case correspondingly.
In the above two exact branes solutions,
$X^\mu_0$ are arbitrary constants, and $\tau(X^a)$ is given in (\ref{tsol1}).
We note that the comparison of the solutions $X^\mu(X^a,\xi^i)$ with
the initial ansatz (\ref{wgac}) for $X^\mu$ shows, that the dependence on
$\La^\mu_0$ has disappeared. We will comment on this later on.

Let us turn to the more complicated case, when the background fields depend on
more than one coordinate $X^a=Y^a(\tau)$. We would like to apply the same
procedure for solving the system of differential equations (\ref{ee}),
(\ref{ec}), as in the simplest case just considered.
To be able to do this, we need to suppose that the metric
$\mathcal{G}_{ab}$ is a diagonal one. Then one can rewrite the effective
equations of motion (\ref{ee}) and the effective constraint (\ref{ec})
in the form
\ba\label{eed} &&\frac{d}{d\tau}\left(\mathcal{G}_{aa}\dot{Y}^a\right)^2 -
\dot{Y}^a\p_a\left(\mathcal{G}_{aa}\mathcal{U}\right)
\\ \nl &&+ \dot{Y}^a\sum_{b\ne a}
\left[\p_a\left(\frac{\mathcal{G}_{aa}}{\mathcal{G}_{bb}}\right)
\left(\mathcal{G}_{bb}\dot{Y}^b\right)^2
- 4\p_{[a}\mathcal{A}_{b]}\mathcal{G}_{aa}\dot{Y}^b\right] = 0,
\\ \label{ecd}
&&\mathcal{G}_{aa}\left(\dot{Y}^a\right)^2
+\sum_{b\ne a}\mathcal{G}_{bb}\left(\dot{Y}^b\right)^2 = \mathcal{U}.\ea

To find solutions of the above equations without choosing particular
background, we fix all coordinates $X^a$ except one. Then the exact
probe brane solution of the equations of motion is given
again by the same expression (\ref{tsol1}) for $\tau\left(X^a\right)$.
In the case when one is using the general ansatz (\ref{wgac}),
the solutions (\ref{xmus1}) and (\ref{xmusd1}) still also hold.

To find solutions depending on more than one coordinate, we have to impose
further conditions on the background fields. Let us show, how a number of
{\it sufficient} conditions, which allow us to reduce the order of the
equations of motion by one, can be obtained.

First of all, we split the index $a$ in such a way that $Y^r$ is one of
the coordinates $Y^a$, and $Y^{\alpha}$ are the others. Then we assume
that the effective 1-form gauge field $\mathcal{A}_a$ can be represented
in the form \ba\label{egfr} \mathcal{A}_a
=(\mathcal{A}_r,\mathcal{A}_\alpha)= (\mathcal{A}_r,\p_\alpha f),\ea i.e.,
it is oriented along the coordinate $Y^r$, and the remaining components
$\mathcal{A}_\alpha$ are pure gauges. Now, the Eq.(\ref{eed}) read
\ba\label{eeda} &&\frac{d}{d\tau}\left(\mathcal{G}_{\alpha\alpha}
\dot{Y}^\alpha\right)^2 -
\dot{Y}^\alpha\p_\alpha\left(\mathcal{G}_{\alpha\alpha}\mathcal{U}\right)
\\ \nl &&+\dot{Y}^\alpha\left[\p_\alpha\left(\frac{\mathcal{G}_{\alpha\alpha}}
{\mathcal{G}_{rr}}\right)\left(\mathcal{G}_{rr}\dot{Y}^r\right)^2
-2\mathcal{G}_{\alpha\alpha}\p_\alpha\left(\mathcal{A}_r-\p_r f\right)
\dot{Y}^r\right]
\\ \nl &&+ \dot{Y}^\alpha\sum_{\beta\ne \alpha}
\p_\alpha\left(\frac{\mathcal{G}_{\alpha\alpha}}{\mathcal{G}_{\beta\beta}}
\right)\left(\mathcal{G}_{\beta\beta}\dot{Y}^\beta\right)^2 = 0,
\\ \label{eedr} &&\frac{d}{d\tau}\left(\mathcal{G}_{rr}\dot{Y}^r\right)^2 -
\dot{Y}^r\p_r\left(\mathcal{G}_{rr}\mathcal{U}\right)
\\ \nl &&+ \dot{Y}^r\sum_{\alpha}
\left[\p_r\left(\frac{\mathcal{G}_{rr}}{\mathcal{G}_{\alpha\alpha}}\right)
\left(\mathcal{G}_{\alpha\alpha}\dot{Y}^\alpha\right)^2
+2\mathcal{G}_{rr}\p_{\alpha}\left(\mathcal{A}_r-\p_r f\right)
\dot{Y}^\alpha\right] = 0.\ea After imposing the conditions \ba\label{ca}
\p_\alpha\left(\frac{\mathcal{G}_{\alpha\alpha}}
{\mathcal{G}_{aa}}\right)=0,
\h\p_\alpha\left(\mathcal{G}_{rr}\dot{Y}^r\right)^2=0,\ea the Eq.
(\ref{eeda}) reduce to \ba\nl
\frac{d}{d\tau}\left(\mathcal{G}_{\alpha\alpha}\dot{Y}^\alpha\right)^2
-\dot{Y}^\alpha\p_\alpha\left\{\mathcal{G}_{\alpha\alpha}\left[\mathcal{U}
+2\left(\mathcal{A}_r-\p_r f\right)\dot{Y}^r\right]\right\}=0,\ea which
are solved by \ba\label{fia}
\left(\mathcal{G}_{\alpha\alpha}\dot{Y}^\alpha\right)^2 =D_{\alpha}
\left(Y^{a\ne\alpha}\right) + \mathcal{G}_{\alpha\alpha}\left[\mathcal{U}
+2\left(\mathcal{A}_r-\p_r f\right)\dot{Y}^r\right]= E_{\alpha}
\left(Y^{\beta}\right)\ge 0,\ea where $D_{\alpha}$, $E_{\alpha}$ are
arbitrary functions of their arguments. \footnote{$E_{\alpha}=E_{\alpha}
\left(Y^{\beta}\right)$ follows from (\ref{cr}).}

To integrate the Eq. (\ref{eedr}), we impose the condition \ba\label{cr}
\p_r\left(\mathcal{G}_{\alpha\alpha}\dot{Y}^\alpha\right)^2=0. \ea After
using the second of the conditions (\ref{ca}), the condition (\ref{cr}),
and the already obtained solution (\ref{fia}), the Eq. (\ref{eedr}) can be
recast in the form \ba\nl
&&\frac{d}{d\tau}\left[\left(\mathcal{G}_{rr}\dot{Y}^r\right)^2
+2\mathcal{G}_{rr}\left(\mathcal{A}_r-\p_r f\right)\dot{Y}^r\right]\\ \nl
&&=\dot{Y}^r\p_r\left\{\mathcal{G}_{rr}\left[(1-n_\alpha)\left(\mathcal{U}
+2\left(\mathcal{A}_r-\p_r f\right)\dot{Y}^r\right)
-\sum_{\alpha}\frac{D_{\alpha} \left(Y^{a\ne\alpha}\right)}
{\mathcal{G}_{\alpha\alpha}}\right]\right\},\ea where $n_\alpha$ is the
number of the coordinates $Y^\alpha$. The solution of this equation,
compatible with (\ref{fia}) and with the effective constraint (\ref{ecd}),
is \ba\label{fir} \left(\mathcal{G}_{rr}\dot{Y}^r\right)^2 =
\mathcal{G}_{rr} \left[(1-n_\alpha)\mathcal{U}
-2n_\alpha\left(\mathcal{A}_r-\p_r f\right)\dot{Y}^r
-\sum_{\alpha}\frac{D_{\alpha} \left(Y^{a\ne\alpha}\right)}
{\mathcal{G}_{\alpha\alpha}}\right]= E_r\left(Y^r\right)\ge 0,\ea where
$E_r$ is again an arbitrary function.

Thus, we succeeded to separate the variables $\dot{Y}^a$ and to obtain the
first integrals (\ref{fia}), (\ref{fir}) for the equations of motion
(\ref{eed}), when the conditions (\ref{egfr}), (\ref{ca}), (\ref{cr}) on
the background are fulfilled. \footnote{An example, when the obtained
sufficient conditions are satisfied, is given by the evolution of a
tensionless brane in Kerr space-time. Moreover, in this case, one is able
to find the orbit $r=r(\theta)$ \cite{Bd01}.} Further progress is
possible, when working with particular background configurations, having
additional symmetries (see, for instance, \cite{KK99}).

\setcounter{equation}{0}
\section{Summary and discussion}

In this paper we addressed the problem of obtaining {\it explicit exact}
solutions for probe branes moving in general string theory backgrounds. We
concentrated our attention to the {\it common} properties of the
$p$-branes and D$p$-branes dynamics and tried to formulate an approach,
which is effective for different embeddings, for arbitrary worldvolume and
space-time dimensions, for different variable background fields, for
tensile and tensionless branes. To achieve this, we first performed an
analysis in Section 2, with the aim to choose brane actions, which are
most appropriate for our purposes.

In Section 3, we formulated the frameworks in which to search for exact
probe branes solutions. The guiding idea is the reduction of the brane
dynamics to a particle-like one. In view of the existing practice, we
first consider the case of {\it static gauge} embedding, which is the
mostly used one in higher dimensions. Then we turn to the more general
case of {\it linear embeddings}, which are appropriate for lower
dimensions too. After that, we consider the branes dynamics by using the
most general ansatz, allowing for its reduction to particle-like one. The
obtained results reveal one common property in all the cases considered.
The {\it effective} equations of motion and one of the constraints, the
{\it effective} constraint, have the {\it same form} independently of the
ansatz used to reduce the $p$-branes or D$p$-branes dynamics. In general,
the effective equations of motion do not coincide with the geodesic ones.
The deviation from the geodesic motion is due to the appearance of {\it
effective} scalar and 1-form gauge potentials. The same scalar potential
arises in the effective constraint.

In the last part of Section 3, we considered the problem of obtaining {\it
explicit exact} solutions of the effective equations of motion and the
effective constraint, without using the explicit structure of the
effective potentials.

In the case when the background fields depend on only one coordinate
$x^a=X^a(\tau)$, we show that these equations can always be integrated and
give the probe brane solution in the form $\tau=\tau(X^a)$, where $\tau$
is the worldvolume temporal parameter. We also give the explicit solutions
for the brane coordinates $X^\mu$ in the form $X^\mu=X^\mu(X^a,\xi^i)$.
They are nontrivial when one uses the most general ansatz (\ref{wgac}).
\footnote{Let us remind that $x^\mu$ are the coordinates, on which the
background fields do not depend.}

In the case when the background fields depend on more than one coordinate,
and we fix all brane coordinates $X^a$ except one, the exact
solutions are given by the same expressions as in the case just considered, if
the metric $\mathcal{G}_{ab}$ is a diagonal one. In this way, we have realized
the possibility to obtain probe brane solutions as functions of {\it every}
single one coordinate, on which the background depends. In the case when none
of the brane coordinates is kept fixed, we were able to find {\it sufficient}
conditions, which ensure the separation of the variables
$\dot{X}^a=\dot{Y}^a(\tau)$. As a result, we have found the manifest
expressions for $n_a$ first integrals of the equations of motion, where
$n_a$ is the number of the brane coordinates $Y^a$.

In obtaining the solutions described above, it was not taken into account that
some restrictions on them can arise, depending on the ansatz used and on the
type of the branes considered. As far as we are interested here in the
{\it common} properties of the probe branes dynamics, we will not
make an exhaustive investigation of all possible peculiarities, which can arise
in different particular cases. Nevertheless, we will consider some
specific properties, characterizing the dynamics of the different type
of branes for different embeddings.

We note that in static gauge, the brane coordinates $X^a$ figuring in our
solutions, are spatial ones. This is so, because in this gauge the background
temporal coordinate, on which the background fields can depend, is
identified with the worldvolume time $\tau$.

The solutions $X^\mu(X^a,\xi^i)$, given by (\ref{xmus1}) for the $p$-brane and
by (\ref{xmusd1}) for the D$p$-brane, depend on the worldvolume parameters
$(\tau,\xi^i)$ through the specific combination
$\La^\mu_i\left(\la^i\tau+\xi^i\right)$. It is interesting to understand if
its origin has some physical meaning. To this end, let us consider the
$p$-branes equations of motion (\ref{pbem}) and constraints (\ref{pbic0}),
(\ref{pbicj}) in the tensionless limit $T_p\to 0$, when they take the form
\ba\nl &&g_{LN}\Di\Dj X^N+\Gamma_{L,MN}\Di X^M \Dj X^N = 0 ,\\ \nl
&&g_{MN}
\left(\p_0-\lambda^i\p_i\right) X^M\left(\p_0-\lambda^j\p_j\right)X^N=0,\\ \nl
&&g_{MN}\left(\p_0-\lambda^i\p_i\right) X^M\p_j X^N=0.\ea
It is easy to check that in $D$-dimensional space-time, any $D$ arbitrary
functions of the type $F^M=F^M\left(\la^i\tau+\xi^i\right)$ solve this system
of partial differential equations. Hence, the linear part of the {\it tensile}
$p$-brane and D$p$-brane solutions (\ref{xmus1}) and (\ref{xmusd1}), is a
background independent solution of the {\it tensionless} $p$-brane equations
of motion and constraints.

Let us point out here that by construction, the actions used in our
considerations allow for taking the tensionless limit $T_p\to 0$
($T_{Dp}\to 0$). Moreover, from the explicit form of the obtained exact probe
branes solutions it is clear that the opposite limit $T_p\to\infty$
($T_{Dp}\to\infty$) can be also taken.

We have obtained solutions of the probe branes equations of motion and one
of the constrains, which have the same form for all of the considered cases.
Now, let us see how we can satisfy the other constraints present in the theory.
These are $p$ constraints, obtained by varying the corresponding actions with
respect to the Lagrange multipliers $\la^i$. For the D$p$-brane, we have $p$
additional constraints, obtained by varying the action with respect to the
Lagrange multipliers $\kappa^i$. Actually, the constraints generated by the
$\la^i$-multipliers are satisfied. Due to the conservation of the corresponding
momenta, they just restrict the number of the {\it independent} parameters
present in the solutions. The only exception is the $p$-brane in static gauge
case, where the momenta $P_i^{SG}$ must be zero. Let us give an example how
the problem can be resolved in a particular situation, which is nevertheless
general enough. Let the background metric along the probe $p$-brane be a
diagonal one. Then from (\ref{cis}) and (\ref{gms}) it follows that the
momenta $P_i^{SG}$ will be identically zero, if we work in the gauge
$\la^i=0$. In the general case, and this is also valid for the $\kappa^i$-
generated constraints, we have to insert the obtained solution of the equations
of motion into the unresolved constraints. The result will be a number of
algebraic relations between the background fields. If they are not satisfied
(on the solution) at least for some particular values of the free parameters
in the solution, it would be fair to say that our approach does not work
properly in this case, and some modification is needed.

Finally, let us say a few words about some possible generalizations of the
obtained results.

As is known, the branes charges are restricted up to a sign to be equal to
the branes tensions from the condition for space-time supersymmetry of the
corresponding actions. In our computations, however, the coefficients in
front of the background antisymmetric fields do not play any special role.
That is why, to account for nonsupersymmetric probe branes, it is enough
to make the replacements \ba\nl T_p b_{p+1} \to Q_p b_{p+1},\h T_{Dp}
c_{p+1} \to Q_{Dp} c_{p+1}.\ea

In our D$p$-brane action (\ref{oda}), we have included only the leading
Wess-Zumino term of the possible D$p$-brane couplings. It is easy to see that
our results can be generalized to include other interaction terms just by the
replacement
\ba\nl  c_{p+1} \to c_{p+1}+c_{p-1}\wedge b_2 + \ldots .\ea
This is a consequence of the fact that we do not used the explicit form of the
background field $c_{M_0\ldots M_p}$. We have used only its antisymmetry and
its independence on part of the background coordinates.

\vspace*{.5cm}
{\bf Acknowledgments}
\vspace*{.2cm}

The author would like to acknowledge the hospitality of the ICTP-Trieste,
where this investigation has been done.
This work is supported by the Abdus Salam International Center for
Theoretical Physics, Trieste, Italy, and by a Shoumen University grant under
contract {\it No.005/2002}.

\end{document}